\documentclass[11pt]{article}
\usepackage{amsmath,amsfonts,epsfig}


\newtheorem{theorem}{Theorem}
\newcounter{lemma}
\newtheorem{lemma}{Lemma}

\newtheorem{claim}{Claim}

\newtheorem{definition}{Definition}
\newtheorem{observation}{Observation}

\newenvironment{proof}{\noindent{\bf Proof:  }}
   {\hfill\rule{2mm}{2mm}}

\newcommand{\ket}[1]{|#1\rangle}

\newcommand{\om}[2]{\omega_{#1}^{#2}}
\newcommand{\ignore}[1]{}
\newcommand{\balpha}{\boldsymbol\alpha}
\newcommand{\bbeta}{\boldsymbol\beta}
\newcommand{\bgamma}{\boldsymbol\gamma}
\newcommand{\bzeta}{\boldsymbol\zeta}
\newcommand{\bdelta}{\boldsymbol\delta}
 
\newcommand{\tens}[1]{\prod [{#1_i}]}
\newcommand{\ft}[2]{\bigotimes_{0\leq i\leq #2}\mbox FT_{#1_i}}

\author{Lisa Hales
\thanks{Group in Logic and the Methodology of Science, University of California at Berkeley, lisah@math.berkeley.edu}
	\and
	Sean Hallgren
\thanks{Computer Science Division, University of California at Berkeley, hallgren@cs.berkeley.edu}
}

\title{Sampling Fourier Transforms on Different Domains
\\(Preliminary Version)
}
\begin{document}
\bibliographystyle{alpha}

\maketitle
\begin{abstract}
We isolate and generalize a technique implicit in many quantum
algorithms, including Shor's algorithms for factoring and discrete
log. In particular, we show that the distribution sampled after a
Fourier transform over ${\mathbb Z}_p$ can be efficiently approximated by
transforming over ${\mathbb Z}_q$ for any q in a large range. Our result
places no restrictions on the superposition to be transformed,
generalizing the result implicit in Shor which applies only to
periodic superpositions. In addition, our proof easily generalizes to
multi-dimensional transforms for any constant number of dimensions.

\end{abstract}

\section{Introduction}

One of the main applications of the fourier transform in quantum
computing is finding a hidden subgroup of a finite abelian group.
Specifically, we are given a finite abelian group $G$ and a function 
$f$ defined on $G$ that is constant and distinct on the cosets of
some unknown subgroup $H$, which we must reconstruct. 

The quantum algorithms solving this problem share a simple conceptual
basis.  Ideally, the machine is put into a uniform superposition of the
elements of some coset of $H$.  Then a fourier transform is performed,
resulting in a uniform superposition on the quotient group $G/H$.  The
subgroup $H$ can be reconstructed after sampling this distribution.
There have been many papers addressing special cases of this problem,
including \cite{Simon94}, \cite{Shor1997}, and \cite{BonehL1995}.
These papers show how to recover the period of a periodic function
defined on ${\mathbb Z}$, in other words they address the case where $H$
and $G$ are cyclic.  \cite{Kitaev1995} solves a more general case,
called the abelian stabilizer problem.

There is also great interest in extending these ideas to non-abelian
groups, in part because the problem of graph isomorphism is reducible
to finding a hidden subgroup in $S_n$.  \cite{Beals1997} shows how
compute a quantum fourier transform of a non-abelian group, but it is
not known how to use this to find a hidden subgroup.  In
\cite{EttingerH1998} an algorithm is given for finding the hidden
subgroup of the dihedral group of order $2N$ that takes exponential
time but has only polynomial query complexity.

In the abelian case, despite the simplicity of the conceptual
framework, technical difficulties arise because it may be impossible
to construct the desired initial superposition or to efficiently
transform over the correct group, or to transform over the correct
group at all if it is not given.  In \cite{Simon94} it is possible to
transform over the group exactly, but these problems arise in
\cite{Shor1997} and \cite{BonehL1995}.  In the case of Shor's discrete
log algorithm, the correct group is known, but can only be efficiently
transformed over if it is ${\mathbb Z}_q$ for a smooth integer $q$.
\cite{Kitaev1995} gives an algorithm for fourier transforming over any
abelian group.  However, in the case of factoring, the ideal domain
(that is, the group) is not even known: not only can the transform not
be performed, but the exact input superposition cannot be constructed.

In \cite{Shor1997} and \cite{BonehL1995} these difficulties are
resolved by transforming over smooth integers satisfying certain
conditions, and providing technical arguments to show that the desired
information can still be reconstructed.  Unfortunately, these
arguments seem particular to each algorithm and obscure the simple
conceptual framework discussed above.

This paper unifies and generalizes these results.  In particular, we
prove that the distribution sampled after a Fourier transform over
${\mathbb Z}_p$ can be efficiently approximated by transforming over
${\mathbb Z}_q$ for any q in a large range. In addition, our proof
easily generalizes to multi-dimensional transforms for any constant
number of dimensions.  This generalizes the previous work by removing
any restrictions on the input distribution (such as periodicity) and
unifying the proofs given for different dimensional transforms.  From
previous work it was not clear that the approach of transforming over
a larger domain would always give the same points as the original set.
Here we show that it does, and we work out the details one once and
for all.  Our result is in fact a mathematical property of the quantum
fourier transform, which makes it easier to design algorithms.  This
also gives an alternative to Kitaev's algorithm.  Instead of using a
more complicated quantum algorithm, the fourier transform is over a
large enough, but otherwise arbitrary, domain.  This would make it
easy to, for example, always transform over a power of 2, while the
conceptual anaylsis requires some other domain.  Also, when the exact
underlying group is not known, as is the case in factoring, algorithms
can still be designed as if it were.

In summary, the following papers discuss computing fourier transforms
efficiently.  \cite{Shor1997} shows how to transform over smooth
numbers.  \cite{Cleve1994} extends this to the case where the prime
factors are not unique but still small.  \cite{Kitaev1995} shows how
to transform over any integer to within any epsilon.
\cite{Coppersmith1994} and \cite{Barenco-et-al-96} show how to
approximate the transform over the same integer by leaving out some
gates.  \cite{Beals1997} shows how to transform over the symmetric
group.  \cite{Dartmouth:TR96-281} gives classical algorithms for
computing the fast fourier transform of functions defined on finite
groups.  \cite{Hoyer1997} gives quantum networks for computing unitary
matrices that can be factored in the right way.

\section{Definitions and Main Theorem }
We will use the following notation throughout our discussion:

\begin{itemize}
\item $\balpha$ is a fixed input superposition:
$\balpha = \sum_{i=0}^{p-1} \alpha_i \ket{i}$

\item $\bbeta$ is the fourier transform of $\balpha$ over domain $p$:
$\bbeta = \sum_{i=0}^{p-1} \beta_i \ket{i} = \mbox{FT}_p (\balpha)$.

\ignore{=\mbox{FT}_p \left(\sum_{i=0}^{p-1} \alpha_i \ket{i} \right)= 
\sum_{c=0}^{p-1} \left( \frac{1}{\sqrt{p}} \sum_{i=0}^{p-1} \om{p}{ic} 
\alpha_i \right) \ket{c}.}

\item $\bgamma$ is the fourier transform of $\balpha$ over domain $q$,
$q>p$:
$\bgamma = \sum_{i=0}^{q-1} \gamma_i \ket{i}w = \mbox{FT}_q (\balpha)$.

\ignore{\mbox{FT}_q \left(\sum_{i=0}^{p-1} \alpha_i \ket{i} \right)= 
\sum_{c=0}^{q-1} \left( \frac{1}{\sqrt{q}} \sum_{i=0}^{p-1} \om{q}{ic} 
\alpha_i \right) \ket{c}.}

\end{itemize}

Figures 1-3 give a simple example of these definitions:

\vspace{.2in}

\begin{figure}[hbt]
  \epsfig{file=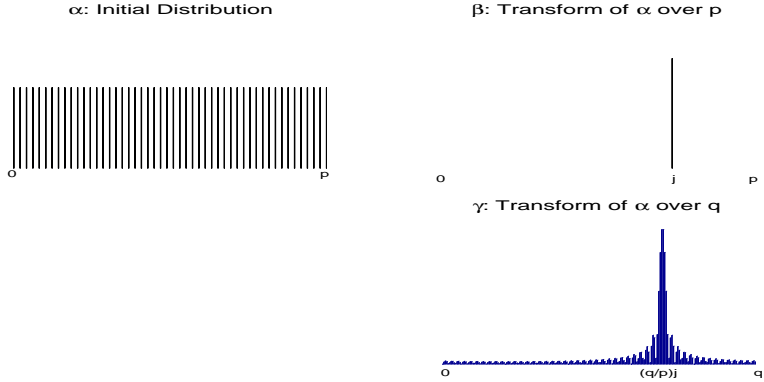,width=4in,height=2in}
  \caption{ $\alpha$, Figure 2: $\beta$, and Figure 3: $\gamma$ }
  \label{fig:delta}
\end{figure}

\vspace{.2in}

Notice that the amplitude at $j$ in figure 2 is centered in figure 3 at
the integers closest to $\frac{q}{p}j$. For this reason the next
definition
will also be useful:

\begin{itemize}
\item For a given index $i$, let $i'$ denote $\lfloor \frac
{q}{p}i\rfloor$.
  If $S\subseteq [p]$ is a set of indices, then
$S'\subseteq [q]$ is the set $\{\lfloor \frac{q}{p}s \rfloor | s\in
S\}$.

\item For $S \subseteq [p]$ and $\bzeta$ a vector of length p, let
$\bzeta_S$ be the vector satisfying 
$(\bzeta_S)_i=(\bzeta)_i$ for all $i\in S$ and $(\bzeta_S)_i=0$
otherwise.

\item The $l_1$ norm of a vector $\bzeta$, denoted $\|\bzeta\|_1$, is 
$\sum_{i=0}^{\dim(\bzeta)-1} |\zeta_i|$.  Likewise the $l_2$ norm 
of a vector $\bzeta$, denoted $\|\bzeta\|_2$, is $
\sqrt{\sum_{i=0}^{\dim(\bzeta)-1}{|\zeta_i|^2}}$.
\end{itemize}

Finally, we need to define the following two distributions:
\begin{itemize}
\item ${\cal D}_{\bbeta}$ is the distribution on $[p]$ induced by
observing the superposition $\bbeta$, i.e.
${\cal D}_{\bbeta} (i)= |\beta_i|^2$.

\item${\cal D}_{\bgamma}$ is the distribution on $[p]$ given
by ${\cal D}_{\bgamma}(i)=\frac{|\bgamma_{i'}|^2}
{\sum_{i\in [p]}
|\gamma_{i'}|^2}=\frac{|\bgamma_{i'}|^2}{\|\bgamma_{[p]'}\|^2}$. 
This is the distribution
on $[p]$ induced by observing the superposition 
$\bgamma$, and outputting $i$ if the observation
is of the form $i'$ for some $i\in [p]$.  Notice that 
if $q$ is a polynomial multiple of $p$ then we will
see points of the form $i'$ with significant probability
and can round to find $i$.  Thus this distribution can be reconstructed
by sampling $\bgamma$.
    
\end{itemize}

We can now state our main theorem, which says that the distribution
sampled after transforming over ${\mathbb Z_p}$ is close to a
distribution which we can efficiently reconstruct by transforming over
${\mathbb Z_q}$ for $q$ a polynomial multiple of $p$.

\begin{theorem}\label{thm:main}
Let $p=O(2^{n^k})$ for some $k$.  Then for any polynomial $s(n)$,
there is a polynomial $t(n)$ such that whenever $q\geq t(n)p$,
$$\|{\cal D}_{\bbeta}-{\cal D}_{\bgamma}\|_1\leq\frac{1}{s(n)}.$$
\end{theorem}

\section{Applications}

Theorem \ref{thm:main} simplifies proofs using fourier transforms.
First we will indicate how to apply it in general and then we will give
some specific applications.

\subsection{General Application}

A general approach to using the fourier transform is as follows:
\begin{itemize}
\item Show that some value $p$ exists such that when transforming over
$p$, and sampling the resulting distribution, we see some set $S$ 
with at least 1/poly probability.
\item Invoke the theorem for some $q$ which
is a polynomial multiple larger than $p$ and which we can 
find and easily transform over, thereby reconstructing $S$.
\end{itemize}

Note that we place no requirements (such as periodicity) on the input distribution. 

\subsection{An Application}

As an example of the application of our theorem, we reprove the
following result of Shor:

\begin{theorem}\label{shthm}(Shor)
Suppose the function $h:{\mathbb Z}\rightarrow{\mathbb Z}x$ is
periodic with period $r$, one-to-one on its fundamental period, and
efficiently computable.  Then in random quantum polynomial time in
$n=\log r$ it is possible to recover $r$.
\end{theorem}

Assume $h$ is as above.  Suppose we could set up the superposition
$\frac{1}{\sqrt{tr}}\sum_{i=0}^{tr}\ket{i, h(i)}$, transform over
$tr$, and sample.  Then we would see $(jt,b)$ with probability

\begin{eqnarray*}
\left|\frac{1}{tr}\sum_{i,h(i)=b}\omega_{tr}^{ijt}\right|^2
&=&\left|\frac{1}{tr}\sum_{k=0}^{t-1}
\omega_{tr}^{(i_0+kr)jt}\right|^2
=\left|\frac{1}{r}\right|^2=\frac{1}{r^2}
\end{eqnarray*}
where $i_0$ satisfies $h(i_0)=b$ and $0\leq i_0<r$.  To reconstruct
the order $r$ we will need to sample $jt$ for $j$ relatively prime to
$r$. The number of such $j$ is $\Phi(r)$, the number of distinct $b$
is $r$.  Thus the probability of seeing a pair $(jt, b)$ with $j$
relatively prime to $r$ is $\frac{r\Phi (r)}{r^2}=\frac{\Phi(r)}{r}$.
Since by a classical result in number theory
$\frac{\Phi(r)}{r}>\frac{k}{\log\log r}$, this probability is at least
$\frac{k}{\log n}$ for some constant $k$.

If we find $jt$ for $j$ relatively prime to $r$ we can compute
$gcd(jt, tr)$ and $\frac{tr}{gcd(jt, tr)}=r$.  Since we can check to
make sure that this is actually the period, using the fact that $h$ is
one-to-one on its fundamental domain, we can keep sampling until we
see a pair of this form, which will happen with high probability
within $O(\log(n))$ repetitions.

Unfortunately, since we do not know $r$, we can neither set up the
desired input superposition nor transform over the desired
domain. Assume for a moment that we could set up the input
superposition $\frac{1}{\sqrt{tr}}\sum_{i=0}^{tr}\ket{i, h(i)}$ for
some $t>r$.  By our theorem, with $s(n)>2\log n$, there is a
polynomial $t(n)$ so that if we transform over a smooth $q$ such that
$t(n)tr<q<2t(n)tr$, then we will see an element of the form
$s_j=\lfloor\frac{q}{tr}jt\rfloor$ with $j$ relatively prime to $r$
with probablity at least $\frac{1}{2t(n)\log n}$. Since
$$\left|s_j-\frac{qj}{r}\right|\leq 1\mbox{, we have }
   \left|\frac{s_j}{q}-\frac{j}{r}\right|\leq \frac{1}{q}.$$ 
Using the fact that $q>r^2$, by rounding $\frac{s_j}{q}$ to the
nearest fraction with denominator less than $\sqrt{q}$, we will find
$\frac{j}{r}$ and thus recover $r$.  We can construct such a $q$ using
the standard method of multiplying together succesively larger primes
until we are in the correct range.

Finally, we must address the fact that we cannot actually construct
the input superposition $\frac{1}{\sqrt{tr}}\sum_{i=0}^{tr}\ket{i,
h(i)}$.  But this problem is easily solved--we will construct a
superposition which is exponentially close to the desired one.  We can
assume without loss of generality that we have an upper bound on $r'$
on $r$ such that $r<r'<2r$.  (If not we can initially set $r'=1$ then
repeatedly run our algorithm, each time doubling our previous guess of
$r'$.) We can easily set up the superposition
$\frac{1}{\sqrt{p}}\sum_{i=0}^{p}\ket{i, h(i)}$ where $p$ is a smooth
number such that $(r')^2<p<2(r')^2$.  This superposition is
exponentially close to $\frac{1}{\sqrt{tr}}\sum_{i=0}^{tr}\ket{i,
h(i)}$ where $tr$ is the multiple of $r$ nearest $p$.

\subsection{An Application}

As a second example of the application of our theorem, we reprove a
result of Boneh and Lipton.  Following their terminology, we say that
the periodic function $h$ has {\bf order} $m$ provided that no more
than $m$ elements in the fundamental domain have the same image under
$h$. Also, a function $f:{\mathbb Z}^2\rightarrow{\mathbb Z}$ has {\bf
hidden linear structure over $q$} provided there is an integer
$\alpha$ and a function $h:{\mathbb Z}\rightarrow{\mathbb Z}$ with
period $q$ such that $f(x,y)=x+\alpha y$.

\begin{theorem}\label{BLthm}(Boneh-Lipton)
Suppose the function $f$ has hidden linear structure over q.  Let $r$
be the smallest positive period of the underlying $h$ and assume $h$
has order at most $m$, where $m$ satisfies the following two
conditions:
\begin{enumerate}
\item Let $n=\log r$, then $m$ is at most $n^{O(1)}$.
\item Let $p$ be the smallest prime divisor of $r$; then $m<p$.
\end{enumerate}
Then, assuming $q$ and $m$ are known and $f$ is efficiently
computable, in random quantum polynomial time in $n$ it is possible to
recover the period $\alpha$.
\end{theorem}
The two conditions on $m$ are required so that the output of the
algorithm can be tested for correctness.

We first need the following lemma.  By using our theorem we are able
to make do with this weakened version of the lemma found in
Boneh-Lipton and considerably simplify the proof.

\begin{lemma}
For any integers $b_1,\dots,b_m$, there are at least $r/m$ elements
$x\in[r]$ satisfying $$\left|\sum_{i=1}^{m}\omega_r^{xb_i}\right|\geq
1/2.$$
\end{lemma}
\begin{proof}(of Lemma)
Note that $\mbox{FT}_r\left(\sum_{i=1}^m\frac{1}{\sqrt{m}}
\ket{b_i\mod r}\right)=
\sum_{j=0}^{r-1}\frac{1}{\sqrt{r}}\left(\frac{1}{\sqrt{m}}
\sum_{i=1}^{m}\omega_r^{xb_i}\ket{j}\right)$.  Thus the number of $x$
satisfying the condition of the theorem is the same as the number of
$x$ with amplitude at least $\frac{1}{2\sqrt{rm}}$ after this
transform.  Suppose there are at most $t$ such $x$'s. Note that the
maximal amplitude after this transform is $\sqrt{\frac{m}{r}}$, thus

$$ 1 \leq t\left(\sqrt{\frac{m}{r}}\right)^2 
+ (r-t)\left(\frac{1}{2\sqrt{rm}}\right)^2$$
 
which implies $$t\geq\frac{4rm-r}{4m^2-1}\geq\frac{r}{m},$$ as desired.
\end{proof}
 
\begin{proof}(of Theorem)
We first  set up the superposition
$\frac{1}{r}\sum_{x_1, x_2}\ket{x_1, x_2}$, then compute $f$,
yielding $$\frac{1}{r}\sum_{x_1, x_2}\ket{x_1, x_2, f(x_1,x_2)}.$$ 

Suppose we could then transform over ${\mathbb Z}_r\times{\mathbb
Z}_r$, sending for each $i\in \{1,2\}$, $x_i$ to $y_i$ with amplitude
$\frac{1}{\sqrt{r}} \omega_r^{x_iy_i}$. Then we would see state
$\ket{y_1,y_2,b}$ with probability
\begin{eqnarray*}
\left|\frac{1}{r^2}\sum_{x_1,x_2:f(x_1,x_2)=b}\omega_r^{x_1y_1+x_2y_2}
  \right|^2
&=&\left|\frac{1}{r^2}\sum_{t:f(t)=b}\sum_{x_2}\omega_r^{(t-\alpha
x_2)y_1+x_2y_2}\right|^2\\
&=&\left|\frac{1}{r^2}\sum_{t:f(t)=b}\omega_r^{ty_1}\sum_{x_2} 
  \omega_r^{x_2(y_2-\alpha y_1)}\right|^2\\
\end{eqnarray*}

Thus if $y_2\equiv\alpha y_1\mod r$ and
$\left|\sum_{t:f(t)=b}\omega_r^{ty_1}\right|^2>\frac{1}{4}$, we will
see $\ket{y_1,y_2,b}$ with probability at least $\frac{1}{4r^2}$.
There are at least $r/m$ distinct $b$'s, and, by our lemma, for each
$b$ there are at least $r/m$ $y_1$'s satisfying the above condition.
Thus we will see a triple of the form $\ket{y,\alpha y\mod r, b}$ with
probability at least $\frac{1}{m^2}$.

Since we cannot necessarily transform over ${\mathbb
Z}_r\times{\mathbb Z}_r$, we now use the two-dimensional version of
our theorem with $s(n)=\frac{1}{2m^2}$ to say that there exists a
polynomial $t(n)$ so that if we transform over ${\mathbb
Z}_{q}\times{\mathbb Z}_{q}$ where $t(n)r<q<2t(n)r$ we will see
triples of the form
$\ket{\lfloor\frac{qy}{r}\rfloor,\lfloor\frac{q\alpha y}{r}\rfloor,
b}$, with probability at least $\frac{1}{2m^2t(n)}$.

With such a triple in hand we can reconstruct a non trivial divisor of
$\alpha$.  First we find $\frac{y}{r}$ by rounding
$\frac{\lfloor\frac{qy}{r}\rfloor}{q}$ to the nearest fraction with
denominator $r$.  Then we do the same for $\frac{\alpha y\mod
r}{r}$. At this point we can proceed as outlined in \cite{BonehL1995}.

Furthermore, as in \cite{BonehL1995}, we can check to make sure that
the triple sampled is of the above form and thus use recursion to
solve our problem.
\end{proof}

\section{Proof of Main Theorem}
\subsection{Outline}

Recall that $\bbeta=\mbox{FT}_p(\balpha)$ and $\bgamma=\mbox{FT}_q(\balpha)$
for some fixed superposition $\balpha$.

The main goal of the proof is to show that if $q>t(n)p$ then for any
set $S$ such that ${\cal D}_{\bbeta} (S)$ is nonnegligible, ${\cal
D}_{\bgamma} (S)$ is approximately ${\cal D}_{\bbeta} (S)$.  The closeness
of the two distributions follows easily from this fact and is proved
in section \ref{fourthsub}.

The central idea in the proof is to show the relationship between
arbitrary $\bbeta$ and the resulting $\bgamma$ by first analyzing the
case in which $\bbeta$ is a $\delta$-function, i.e, $\beta=\ket{j}$
for some $j\in [p]$.  In this case $\bgamma$ is ``almost'' a
$\delta$-function, i.e., its amplitude is highly concentrated at
$j'=\lfloor (q/p)j \rfloor$, and we can derive a lower bound on the
amplitude located at $j'$ and an upper bound on the amplitude located
at any other primed index.  These bounds are stated in Claim 1.  We
then extend the analysis from the case of $\delta$-functions to
arbitrary $\bbeta$ using linearity of the transform.  This is the
content of section 4.2.

There is a complication in proving the theorem however.  To use the
bounds derived from the $\delta$-functions, the amplitudes in $S$ must
be approximately equal.  Loosely speaking, in section \ref{secondsub}
we show closeness of ${\cal D}_{\bgamma} (S)$ and ${\cal D}_{\bbeta}
(S)$ when the set satisfies this property (lemma 1), and in section
\ref{thirdsub} we split an arbitrary set $S$ into subsets with
approximately equal amplitudes, apply the previous result to each
subset, and combine the results (lemma 2).

\subsection{Claim \ref{firstclaim}}\label{firstsub}

To prove Lemma \ref{firstlemma} we need to establish
a relationship between the entries of $\bbeta$ and the
primed entries of $\bgamma$. In particular, we would like
to have a lower bound on $|\gamma_{s'}|$ in terms of $|\beta_s|$.
Unfortunately, in general, $|\gamma_{s'}|$ depends on all the entries of
$\bbeta$, not just on $\beta_s$.
However, if $\bbeta$ is a $\delta$-function, i.e. $\bbeta=\ket{j}$
for some $j$, then, all other entries being $0$, $\bgamma_{j'}$ does
depend only on
$\bbeta_j$. Furthermore, we can use this case to derive the general 
relationship between $|\gamma_{s'}|$ and the entries of $\bbeta$.  Thus
we first make the following claim, whose proof can be found in
the appendix:

\begin{claim} 
\label{firstclaim}
Let
$\sum_{i=0}^{q-1}\eta_i\ket{i}= \mbox{FT}_q \mbox{FT}^{-1}_p(\ket{j})
=\mbox{FT}_q\left(\sum_{i=0}^{p-1}\frac{1}
{\sqrt{p}}\om{p}{-ij}\ket{i}\right)$ for some $q>2p$ and $j\in [p]$.
Then the following bounds hold:
\begin{itemize}
\item $|\eta_{j'}| \geq \sqrt{\frac{p}{q}} \left( 1-
20\frac{p^2}{q^2} \right)$
\item For $k\neq j$, 
$|\eta_{k'}| \leq \sqrt{\frac{p}{q}} \frac{2}{|k-j|_p}\frac{p}{q}$
\end{itemize}

where $|x|_p = \left\{ \begin{array}{ll}
		x\bmod p &  \mbox {if $0\leq x\bmod p\leq p/2$}\\
                -x\bmod p & \mbox {otherwise}
		\end{array}
		\right.  $

\end{claim}

This claim is again illustrated in Figures 1-3.  It says that if one
looks where the delta function goes if it is inverse transformed over
$p$ and transformed over $q$, at the spot $j'$ there will still be a
large amplitude, and at any other $k'$, the curves falls off at about
1 over the distance from $j'$.

We can use our claim to derive a lower bound on $|\bgamma_{j'}|$ given an
arbitrary $\bbeta$. We view $\bbeta$ as a complex-weighted sum of
$\delta$-functions, the 
$\delta$-function at $i$ receiving weight $\beta_i$. As in the claim,
the amplitude $\gamma_{j'}$ will receive a contribution of at least
$|\beta_j|\sqrt{\frac{p}{q}} \left( 1-
20\frac{p^2}{q^2} \right)$ from the weighted $\delta$-function at $j$.
On the other hand it will
also receive a contribution of at most $|\beta_k|\sqrt{\frac{p}{q}}
\frac{2}{|k-j|_p}\frac{p}{q}$ from the $\delta$-function
at $k$ for each $k\neq j$. In the worst case these two types of
contributions will be pointed in opposite directions, leading to a lower
bound:

$$|\gamma_{j'}| \geq|\beta_j|\sqrt{\frac{p}{q}} \left( 1-
20\frac{p^2}{q^2} \right)-\sum_{k\neq j}|\beta_k|\sqrt{\frac{p}{q}}
\frac{2}{|k-j|_p}\frac{p}{q}.$$

More formally, by linearity of the transform, $\balpha=\mbox{FT}^{-1}_p 
(\bbeta) = \mbox{FT}^{-1}_p \left(\sum_{j=0}^{p-1} \beta_j \ket{j})\right) = 
\sum_{j=0}^{p-1} \beta_j \mbox{FT}^{-1}_p(\ket{j})$, so
$\bgamma=\mbox{FT}_q (\balpha) = \sum_{j=0}^{p-1} \beta_j \mbox{FT}_q
(\mbox{FT}^{-1}_p (\ket{j}))$. 
Thus, for any particular $j$,
we have
$$\gamma_{j'}=\left(\sum_{k=0}^{p-1}\beta_k\mbox{FT}_q\left(\sum_{i=0}^{
p-1}\frac{
1}{\sqrt {p}}\om{p}{-ij}\ket{i}\right)\right)_{j'}=\beta_j \left(
\mbox{FT}_q\left(\sum_{i=0}^{p-1}
\frac{1}{\sqrt {p}}\om{p}{-ij}\ket{i}\right) \right)_{j'} + \sum_{k\neq
j}\beta_k \left( \mbox{FT}_q\left(\sum_{i=0}^{p-1}\frac{1}{\sqrt
{p}}\om{p}{-ik}\ket{i}\right)\right)_{j'}$$
By our claim, then, \begin{equation}\label{firsteqn}
|\gamma_{j'}|\geq |\beta_j|\sqrt{\frac{p}{q}} \left( 1-
20\frac{p^2}{q^2} \right)-\sum_{k\neq j}|\beta_k|\sqrt{\frac{p}{q}}
\frac{2}{|k-j|_p}\frac{p}{q}.
\end{equation}

Since our goal is to establish that $\|\bgamma_{S'}\|_2^2$ is
approximately $\frac{p}{q}\|\bbeta_{S}\|_2^2$, if we could show that,
when q is chosen to be a sufficiently large polynomial multiple of p,
the second of the two terms above is always negligible compared to the
first, we would be done.  Unfortunately, this is not true -- there
will in fact be indices $s$ with $|\beta_s|$ large where this second
term entirely cancels the first.  In particular, this can happen if
there is an index $t$, close enough to $s$ that $\frac{2}{|t-s|_p}$ is
not too small, whose amplitude, $|\beta_t|$, is more than a polynomial
factor larger than $|\beta_s|$. But, there is not enough total
amplitude in the superposition for this to happen at very many points
in $S$. What we will show, then, is that there is a choice of q so
that for a {\bf typical} point in $S$ the second term in is negligible
compared to the first, in  other words, we can bound $$\sum_{s\in S}
\sum_{t\neq s}|\beta_t| \sqrt{\frac{p}{q}}
\frac{2}{|t-s|_p}\frac{p}{q}.$$

The following argument and bound formalize the intuition that there is
not enough total amplitude to wipe out
most points in $S$:

Since,
\begin{eqnarray*}
\sum_{s\in S} \sum_{t\neq s} \frac{2}{|t-s|_p} |\beta_t|
& = &
\sum_{s\in S} \sum_{t\neq s, |\beta_t|\leq 1/\sqrt{|S|}}
\frac{2}{|t-s|_p} 
|\beta_t| +
\sum_{t, |\beta_t| > 1/\sqrt{|S|}} |\beta_t| \sum_{s\in S,\neq t}
\frac{2}{|t-s|_p}  \\
& \leq & \left(\frac{1}{\sqrt{|S|}} \sum_{s\in S} 4 \ln p \right)+\left(
4 \ln p \sum_{t, |\beta_t| > 1/\sqrt{|S|}} |\beta_t|\right) \\
& \leq &
8 \sqrt{|S|} \ln p,
\end{eqnarray*}

we have 
\begin{equation}\label{secondeqn}
\sum_{s\in S} \sum_{t\neq s}|\beta_t|
\sqrt{\frac{p}{q}}\frac{2}{|t-s|_p}\frac{p}{q}\leq
\left(\frac{p}{q}\right)^{3/2}8 \sqrt{|S|} \ln p.
\end{equation}
We will use both the numbered inequalities derived in this section in
our proof of Lemma \ref{firstlemma}.

{\bf Acknowledgements:} We thank Umesh Vazirani for many useful conversations.

\bibliography{samplft}
\section{Appendix}

\subsection{Proof of Lemma \ref{firstlemma}}\label{secondsub}

\begin{definition}
A vector $\bzeta$ is called {\bf $\bdelta$-uniform} if for all $i,j$
such that $\bzeta_i$ and
$\bzeta_j$ are both non-zero, 
$$\delta\leq\frac{|\bzeta_i|}{|\bzeta_j|}\leq\frac{1}{\delta}$$.
\end{definition}

\begin{lemma} \label{firstlemma} Suppose that $\bbeta_S$ is
$\delta$-uniform and $\|\bbeta_S\|_2^2 = c$. Then if
$q>\left(\frac{3200r\ln p}{\delta\sqrt{c}}\right)p$,
$$\|\bgamma_{S'}\|_2^2 \geq
\frac{p}{q}\delta^2
  \left(1-\frac{1}{100r}\right) c.$$
\end{lemma}

\vspace{.2in}
\begin{proof}(of Lemma \ref{firstlemma})

We will lower bound $\|\bgamma_S\|_1$ in terms of $\|\bbeta_S\|_1$.
Then using $\delta$-uniformity $\|\bbeta_S\|_1$ can be lower bounded in
terms of $\|\bbeta_S\|_2^2$. By a simple minimization principle, this
gives a lower bound on $\|\bgamma_S\|_2^2$ in terms of
$\|\bbeta_S\|_2^2$, as desired.

Using Inequality \ref{firsteqn} from the previous section we can derive
the following lower bound on $\|\bgamma_{S'}\|_1$:
\begin{eqnarray*}
\|\bgamma_{S'}\|_1 
&=&\sum_{s\in S}|\gamma_{s'}|\\
& \geq &
\sum_{s\in S} \left(|\beta_s|\sqrt{\frac{p}{q}} \left(
1-20\frac{p^2}{q^2}
\right)
-\sum_{t\neq s}|\beta_t|\sqrt{\frac{p}{q}} \frac{2}{|t-s|_p} 
 \frac{p}{q} \right) \\
& = & 
\sqrt{\frac{p}{q}} \left( \left( 1-20\frac{p^2}{q^2} \right) 
\|\bbeta_{S}\|_1 - 
\frac{p}{q} \sum_{s\in S} \sum_{t\neq s} \frac{2}{|t-s|_p} |\beta_t| 
\right)
\end{eqnarray*}

Because $S$ is $\delta$-uniform, we can derive the following
lower bound on the $l_1$-norm of $\|\bbeta_{S}\|_1 $:

$$\|\bbeta_{S}\|_1 \geq
\frac{1+\delta}{\sqrt{2}\sqrt{1+\delta^2}} \sqrt{|S|} 
  {\|\bbeta_{S}\|_2} \geq
\delta \sqrt{|S|}{\|\bbeta_{S}\|_2},$$

where the first inequality comes from looking at the worst-case scenario
(half the entries of $\bbeta_S$ are of maximal size and the other half
are
of minimal size), and the second is just algebra.

Thus $$\|\bgamma_{S'}\|_1\geq\sqrt{\frac{p}{q}} \left( \left(
1-20\frac{p^2}{q^2} \right) \delta \sqrt{|S|} {\|\bbeta_{S}\|_2} - 
\frac{p}{q} \sum_{s\in S} \sum_{t\neq s} \frac{2}{|t-s|_p} |\beta_t| 
\right).$$
We upper bound the second term in this difference using Inequality
\ref{secondeqn} from the previous section: 
$$\sum_{s\in S} \sum_{t\neq s}|\beta_t|
\sqrt{\frac{p}{q}}\frac{2}{|t-s|_p}\frac{p}{q}\leq
\left(\frac{p}{q}\right)^{3/2}8 \sqrt{|S|} \ln p.$$
Thus,
\begin{eqnarray*}\|\bgamma_{S'}\|_1
& \geq &
\sqrt{\frac{p}{q}} \left( \left( 1-20\frac{p^2}{q^2} \right) \delta 
\sqrt{|S|} {\|\bbeta_{S}\|_2} -
\frac{p}{q}8 \sqrt{|S|} \ln p
\right)\\
& = & 
\sqrt{\frac{p}{q}}\delta \sqrt{|S|}\sqrt{c}\left( \left( 
1-20\frac{p^2}{q^2} \right) - \frac{p}{q} 
\frac{8 \ln p}{\delta\sqrt{c}} \right) 
\end{eqnarray*}
which implies that
$$\|\bgamma_{S'}\|_2^2 
\geq
\frac{p}{q}\delta^2 c\left(1-20\frac{p^2}{q^2}-
\frac{p}{q} \frac{8 \ln p}{\delta\sqrt{c}} \right)^2.$$
Finally, using our assumption that $q>\left(\frac{3200r\ln
p}{\delta\sqrt{c}}\right)p$,
$$\|\bgamma_{S'}\|_2^2 
\geq
\frac{p}{q}\delta^2 c\left(
1-\frac{1}{100r}\right),$$

as desired.
\end{proof}
\subsection{Lemma \ref{secondlemma}}\label{thirdsub}

Using the bound for $\delta$-uniform sets in Lemma \ref{firstlemma}, we
can establish the following bound for arbitrary sets $S$:

\begin{lemma}\label{secondlemma} 

If $\|\bbeta_S\|_2^2 = c$ and $q\geq\left(\frac{6400r\ln
p\sqrt{\ln\frac{c}{|S|100r}}}{\sqrt{c\ln(1-\frac{1}{100r})}}\right)p$,
then
$$\|\bgamma_{S'}\|_2^2 \geq
\frac{p}{q}
  \left(1-\frac{1}{r}\right)c$$
\end{lemma}

\begin{proof}(of Lemma \ref{secondlemma} from Lemma \ref{firstlemma})

Lemma \ref{secondlemma} follows fairly easily from Lemma
\ref{firstlemma}. The idea is to first remove
from $S$ indices corresponding to insignificantly
small amplitudes. Then partition the new $S$ into a collection of
$\delta$-uniform subsets.  We can apply Lemma \ref{firstlemma} to 
each $\delta$-uniform subset of sufficiently large probability, and the 
total probability of the remaining, small $\delta$-uniform subsets is 
insignificant.

First, discard all indices in $s\in S$ with
$|\beta_s|<\sqrt{\frac{c}{100r|S|}}$. Since
we have thrown out at most $|S|$ such indices, we have lost at most
$\frac{c}{100r}$ in probability
and we have $\|\bbeta_S\|_2^2 \geq c\left(1-\frac{1}{100r}\right)$.

Partition $S$ into subsets $$S_i = \{s\in S| \delta^i < |\beta_s| \leq
\delta^{i-1}\}$$ 
for $0<i\leq \log_{1/\delta} \left(\sqrt{\frac{|S| 100r}{c}} \right)$
and $\delta=(1-\frac{1}{100r})$.

In what follows let
$T=\{i:\|\bbeta_{S_i}\|_2^2\geq\frac{c}{log_{1/\delta}\sqrt{\frac{|S|100
r}{c}}}\}$.  Since $$q\geq\left(\frac{6400r\ln
p\sqrt{\ln\frac{c}{|S|100r}}}{\sqrt{c\ln(1-\frac{1}{100r})}}\right)p\geq
\left(\frac{3200r\ln p}{\delta\sqrt{\min_{i\in
T}\|\bbeta_{S_i}\|_2^2}}\right)p,$$ we can apply Lemma \ref{firstlemma}
for each $i\in
T$.

Thus,
\begin{eqnarray*} 
\|\bgamma_{S'}\|_2^2
&=&
\sum_i \|\bgamma_{S'_i}\|_2^2\\
&\geq&
\sum_{i\in T} \|\bgamma_{S'_i}\|_2^2\\
&\geq&
\sum_{i\in T} \frac{p}{q}\delta^2\|\bbeta_{S_i}\|_2^2
\left(1-\frac{1}{100r}\right)\\
&=&
\frac{p}{q}\delta^2\left(\sum_{i\in T}\|\bbeta_{S_i}\|_2^2+\sum_{i\notin
T}\|\bbeta_{S_i}\|_2^2\right)\left(1-\frac{1}{100r}\right)-\frac{p}{q}
\delta^2\sum_{i\notin
T}\|\bbeta_{S_i}\|_2^2\left(1-\frac{1}{100r}\right)\\ 
&=&
\frac{p}{q}\delta^2\left(\|\bbeta_S\|_2^2\right)\left(1-\frac{1}{100r}
\right)-\frac{p}{q}
\delta^2\sum_{i\notin
T}\|\bbeta_{S_i}\|_2^2\left(1-\frac{1}{100r}\right)
\end{eqnarray*}
Since\ $\|\bbeta_S\|_2^2\geq c\left(1-\frac{1}{100r}\right)$ and
$$\sum_{i\notin T}\|\bbeta_{S_i}\|_2^2\leq |T|\max_{i\notin
T}\left(\|\bbeta_{S_i}\|_2^2\right)\leq\log_{1/\delta}
\left(\sqrt{\frac{|S| 100r}{c}}
\right)\frac{c}{log_{1/\delta}\sqrt{\frac{|S|100r}{c}}},$$ we have
\begin{eqnarray*} 
\|\bgamma_{S'}\|_2^2
&\geq&
\frac{p}{q}\delta^2
c\left(1-\frac{1}{100r}\right)^2-\frac{p}{q}\delta^2\log_{1/\delta}
\left(\sqrt{\frac{|S| 100r}{c}}
\right)\frac{c}{log_{1/\delta}\sqrt{\frac{|S|100r}{c}}}\left(1-\frac{1}{
100r}\right)\\
&=&
\frac{p}{q}c\delta^2\left(\left(1-\frac{1}{100r}\right)^2-\left(1-\frac{
1}{100r}\right)\right)\\
&\geq&
\frac{p}{q}c\left(1-\frac{1}{100r}\right)^2\left(\left(1-\frac{1}{100r}
\right)^2-\left(1-\frac{1}{100r}\right)\right)\\
&\geq&
\frac{p}{q}c\left(1-\frac{1}{r}\right),\\
\end{eqnarray*}
as desired.
\end{proof}

\subsection{Proof of Main Theorem from Lemma
\ref{secondlemma}}\label{fourthsub}

Let $p=O(2^{n^k})$ and $s(n)$ be given. Let $t(n) = \frac{6400r\ln
p\sqrt{\ln\frac{c}{|S|100r}}}{\sqrt{c\ln(1-\frac{1}{100r})}}$ with
$r=4s(n)$ and $c=\frac{1}{2s(n)}$.

Let $R=\{i\in [p]:{\cal D}_{\bbeta}(i)-{\cal D}_{\bgamma}(i)\geq 0\}$.  
Since
\begin{eqnarray*}
\|{\cal D}_{\bbeta}-{\cal D}_{\bgamma}\|_1
&=&\sum_{i\in [p]}\left|{\cal D}_{\bbeta}(i)-{\cal
D}_{\bgamma}(i)\right|\\
&=&\sum_{i\in R}\left({\cal D}_{\bbeta}(i)-{\cal D}_{\bgamma}(i)\right)
+\sum_{i\notin R}\left( {\cal D}_{\bgamma}(i)-{\cal
D}_{\bbeta}(i)\right),\\
\end{eqnarray*}
if $\|{\cal D}_{\bbeta}-{\cal D}_{\bgamma}\|_1>
\frac{1}{s(n)}$ then one of the above two sums must be at least
$\frac{1}{2s(n)}$.
Assume that $\sum_{i\in R}\left({\cal D}_{\bbeta}(i)-{\cal
D}_{\bgamma}(i)\right)>\frac{1}{2s(n)}$.  Then
since $\sum_{i\in R} |\beta_i|^2>\frac{1}{2s(n)}$, we can apply Lemma 1
with $r=4s(n)$ and $c=\sum_{i\in R} |\beta_i|^2>\frac{1}{2s(n)}$. Note
also
that $\|\bgamma_{[p]'}\|^2_2\leq\frac{p}{q}$, thus\\
\begin{eqnarray*}
\sum_{i\in R}\left({\cal D}_{\bbeta}(i)-{\cal D}_{\bgamma}(i)\right)
&=&\|\bbeta_R\|_2^2-\frac{\|\bgamma_{R'}\|_2^2}{\|\bgamma_{[p]'}\|_2^2}\\
&\leq&
\|\bbeta_R\|_2^2-\left(\frac{p}{q}
\left(1-\frac{1}{4s(n)}\right)\frac{\|\bbeta_R\|_2^2}{\|\bgamma_{[p]'}\|
_2^2}\right)\\
&\leq& \|\bbeta_R\|_2^2\frac{1}{4s(n)}\\
&\leq& \frac{1}{2s(n)},
\end{eqnarray*}
a contradiction, as desired.

On the other hand, if $\sum_{i\notin R}\left( {\cal
D}_{\bgamma}(i)-{\cal D}_{\bbeta}(i)\right)>\frac{1}{2s(n)}$ then again
applying
lemma 1 and using the fact that
$\|\bgamma_{R'}\|_2^2\leq\frac{p}{q}\|\bbeta_R\|_2^2$,

\begin{eqnarray*}
\sum_{i\notin R}\left( {\cal D}_{\bgamma}(i)-{\cal D}_{\bbeta}(i)\right)
&=&\frac{\|\bgamma_{R'}\|_2^2}{\|\bgamma_{[p]'}\|_2^2}-\|\bbeta_R\|_2^2\\
&\leq&
\frac{\frac{p}{q}\|\bbeta_R\|_2^2}{\frac{p}{q}\left(1-\frac{1}{4s(n)}
\right)\|\bbeta_{[p]}\|_2^2}-\|\bbeta_R\|_2^2\\
&=& \|\bbeta_R\|_2^2\left(\frac{1}{1-\frac{1}{4s(n)}}-1\right)\\
&\leq& \frac{1}{2s(n)},
\end{eqnarray*}
also a contradiction.

\subsection{Proof of Claim} 

{\bf Claim \ref {firstclaim}} Let
$\sum_{i=0}^{q-1}\eta_i\ket{i}=\mbox{FT}_q\left(\sum_{i=0}^{p-1}\frac{1}
{\sqrt{p}}\om{p}{-ij}\ket{i}\right)$ for some $q>2p$ and $j\in [p]$.
Then
the following bounds hold:
\begin{enumerate}
\item $|\eta_{j'}| \geq \sqrt{\frac{p}{q}} \left( 1-
20\frac{p^2}{q^2} \right)$
\item For $k\neq j$, $|\eta_{k'}| \leq \sqrt{\frac{p}{q}}
\frac{2}{|k-j|_p} 
\frac{p}{q}$
\end{enumerate}
where $|x|_p = \left\{ \begin{array}{ll}
		x\bmod p &  \mbox {if $0\leq x\bmod p\leq p/2$}\\
                -x\bmod p & \mbox {otherwise}
		\end{array}
		\right.  $

\begin{proof}
The first bound is established as follows:

For some $\epsilon$ satisfying $0\leq \epsilon <1$, 
\begin{eqnarray*}
\eta_{j'}
&=& \frac{1}{\sqrt{q}} \sum_{i=0}^{p-1} \frac{1}{\sqrt{p}} \om{p}{-ij}
\om{q}{i(jq/p+\epsilon)}\\
&=& \frac{1}{\sqrt{q}} \sum_{i=0}^{p-1} \frac{1}{\sqrt{p}} \om{p}{-ij}
\om{p}{ij}\om{q}{i\epsilon}\\
&=& \sqrt{\frac{p}{q}} \frac{1}{p} \sum_{i=0}^{p-1} \om{q}{i\epsilon}
\end{eqnarray*}

Since $\left| \frac{1}{p} \sum_{i=0}^{p-1} \om{p}{i\epsilon p/q} \right|
\geq
\cos(2\pi \epsilon p/q) \geq 1-\frac{(2\pi \epsilon p/q)^2}{2}\geq
1-20(p/q)^2$, we have
$|\eta_{j'}| \geq \sqrt{\frac{p}{q}} \left( 1- 20\frac{p^2}{q^2}
\right)$, as desired.

The second bound requires the following observation:

\begin{observation} Let $\delta =|x-\lfloor x \rceil|$.  
Then $\left| \frac{1}{p}\sum_{i=0}^{p-1} \om{p}{ix} \right| \leq
\frac{\delta}{|x|_p}$, whenever the latter expression is defined.
\end{observation}

Using this observation we can prove the second bound as follows:

For some $\epsilon$ satisfying $0\leq \epsilon <1$,
\begin{eqnarray*}
\eta_{k'} 
&=& \frac{1}{\sqrt{q}} \sum_{i=0}^{p-1} \frac{1}{\sqrt{p}} \om{p}{-ij}
\om{q}{i(kq/p+\epsilon)}\\
&=& \sqrt{\frac{p}{q}} \frac{1}{p} \sum_{i=0}^{p-1} \om{p}{i(k-j +
\epsilon\frac{p}{q})}
\end{eqnarray*}

Using our observation, with $\delta=\min(\epsilon\frac{p}{q},
1-\epsilon\frac{p}{q})$, and the fact that $q>2p$, we have
 $|\eta_{k'}| \leq
\sqrt{\frac{p}{q}}\frac{\delta}{|k-j+\epsilon\frac{p}{q}|_p}\leq
\sqrt{\frac{p}{q}}\frac{2}{|k-j|_p}\frac{p}{q}$, as desired.
\end{proof}

\begin{proof}(of observation)
Since $\left|\sum_{i=0}^{p-1} \om{p}{ix}\right|=\left|\sum_{i=0}^{p-1}
\om{p}{i|x|_p}\right|$, 
we will bound the latter sum instead. For ease of reading, let $y=|x|_p$
in what follows. Note
that $\delta=|x-\lfloor x \rceil|=|y-\lfloor y \rceil|$.

First we rewrite each vector in the 
sum $\sum_{i=0}^{p-1} \om{p}{iy}$ 
as an integral over an arc of a circle, in particular, we substitute
$\frac{p}{\pi y}\int_{iy-y/2}^{iy+y/2} \om{p}{t} dt$ for $\om{p}{iy}$.
Then 
\begin{eqnarray*}
\left|\sum_{i=0}^{p-1} \om{p}{iy}\right| 
&=& \left|\frac{p}{\pi y} \sum_{i=0}^{p-1} \int_{iy-y/2}^{iy+y/2}
\om{p}{t} dt\right|\\ 
&=& \left|\frac{p}{\pi y} \int_{-y/2}^{(p-1)y+y/2} \om{p}{t} dt
\right|\\
&=& \left|\frac{p}{\pi y} \int_{-y/2}^{yp-y/2} \om{p}{t} dt \right|\\
&=& \left|\frac{p}{\pi y} \int_0^{yp} \om{p}{t} dt \right|\\
&=& \left|\frac{p}{\pi y} \left(\int_0^{\lfloor y \rceil p} \om{p}{t} dt
+
\int_{\lfloor y \rceil p}^{yp} \om{p}{t} dt\right) \right|\\
&=& \left|\frac{p}{\pi y}   \int_{\lfloor y \rceil p}^{yp} 
\om{p}{t} dt\right|\\
&=& \left|\frac{p}{\pi y}   \int_{0}^{\delta p} 
\om{p}{t} dt\right|\\
&=&\frac{p\delta}{y}\\
\end{eqnarray*}
Thus $\left| \frac{1}{p}\sum_{i=0}^{p-1} \om{p}{ix} \right| \leq
\frac{\delta}{|x|_p}$, as desired.

\end{proof}


\subsection{Multiple Dimensions}

A analogous proof can be given in the case of multi-dimensional
Fourier transforms. First we need to define

\begin{itemize}

\item $\bbeta = \sum_{\vec x\in \tens{p}}\beta_{\vec x}\ket{\vec
x}=\ft{p}{k}(\balpha)$ for some superposition $\balpha$,

\item $\bgamma = \sum_{\vec x\in \tens{q}}\gamma_{\vec x}\ket{\vec
x}$ is $\ft{q}{k}(\balpha)$, and

\item $S' = \{(\lfloor \frac{q_1}{p_1}s_1 \rceil ,\lfloor
\frac{q_2}{p_2}s_2 \rceil ,\dots,\lfloor \frac{q_k}{p_k}s_k \rfloor |
(s_1, s_2,\dots, s_k) \in S\}$. Likewise $\vec y'$ satisfies $(\vec
y')_i=\lfloor\frac{q_i}{p_i}(\vec y)_i\rfloor$.

\end{itemize}

Now we can assert the following lemma which is the multidimensional
version of our Lemma 2:

\begin{lemma} \label{multilem}
If $\|\bbeta_S\|_2^2 = c$ for some set $S\subseteq \tens{p}$, and for
all i, $q_i > \left(\frac {2^{k+2}k^{k+1}800r(\ln p)^k
\sqrt{\ln\frac{c}{|S|100r}}}{\sqrt{c\ln(1-\frac{1}{100r})}}\right)p_i$,
then $\|\bgamma_{S'}\|_2^2 \geq (\prod_{i=1}^{i=k} {\frac{p_i}{q_i}})
\left(1-\frac{1}{r}\right) c$.
\end{lemma}

Notice that the quantity in parentheses is a polynomial whenever $k$,
the number of dimensions is constant.  Using this lemma we can prove
the multidimensional version of our theorem precisely as we did in the
one dimensional case.

To prove the above lemma we will need a generalization of our Claim 1.
In what follows let $\mbox{FT}_{\vec p}=\ft{p}{k}$ and
$\mbox{FT}_{\vec q}=\ft{q}{k}$.

\begin{claim} Let $\bzeta$ satisfy $\mbox{FT}_{\vec p}(\bzeta)=\ket{\vec
y}$ for
some $\vec y\in\tens{p}$. Let $\sum_{\vec x\in\tens{q}}\eta_{\vec
x}\ket{\vec x}=\mbox{FT}_{\vec q}\left(\bzeta\right)$ for some $q_i$
such that for all $i$, $q_i>2p_i$. Then the following bounds hold:

\begin{enumerate}
\item $|\eta_{\vec y'}| \geq \prod_{i=1}^{i=k}\sqrt{\frac{p_i}{q_i}}
\left( 1- 20\frac{p_i^2}{q_i^2}\right)$
\item For $\vec z\neq \vec y$, $|\eta_{\vec z'}| \leq
\left(\prod_{i=1}^{i=k}\sqrt{\frac{p_i}{q_i}}\right)\left( \prod_{j,
\vec z_j\neq\vec y_j}\frac{2}{|\vec z_j-\vec
y_j|_p}\frac{p_j}{q_j}\right)$
\end{enumerate}

\end{claim}
This claim, as in the one dimensional case, allows us to give the
following lower bound on $|\bgamma_{\vec x'}|$:

$$|\gamma_{\vec x'}|\geq |\beta_{\vec
x}|\prod_{i=1}^{i=k}\sqrt{\frac{p_i}{q_i}} \left( 1-
20\frac{p_i^2}{q_i^2}\right)-\sum_{\vec z\neq \vec x}|\beta_{\vec
z}|\left(\prod_{i=1}^{i=k}\sqrt{\frac{p_i}{q_i}}\right)\left(
\prod_{j,
\vec z_j\neq\vec x_j}\frac{2}{|\vec z_j-\vec
x_j|_p}\frac{p_j}{q_j}\right).$$

As in the proof of the one dimensional case, we will need to upper
bound the following quantity:

$$\sum_{\vec x\in S}\sum_{\vec z\neq \vec x}|\beta_{\vec z}|\left(
\prod_{j, \vec z_j\neq\vec x_j}\frac{2}{|\vec z_j-\vec
x_j|_p}\frac{p_j}{q_j}\right).$$

Using an argument which is analogous to the one-dimensional case we
get a bound of
$$2^{k+2}k^{k+1}\sqrt{|S|}(\ln p)^k\min_i\{\frac{p_i}{q_i}\}$$

Using this bound we can carry out the rest of the proof precisely as
in the one dimensional case to get the factors specified in Lemma
\ref{multilem}.


\end{document}